\documentclass[preprint]{aastex}

\slugcomment{}

\shorttitle{Normal, Dim and Dwarf galaxies}
\shortauthors{Driver}

\begin{document}


\title{The contribution of normal, dim and dwarf
    galaxies to the local luminosity density}


\author{S. P. Driver}
\affil{School of Physics and Astronomy, 
University of St Andrews, North Haugh, St Andrews, Fife, KY16 9SS, SCOTLAND}



\begin{abstract}
From the Hubble Deep Field catalog presented in \citet{spd98}
we derive the local ($0.3 < z < 0.5$) {\it bivariate brightness distribution} 
(BBD) of field galaxies within a 326 Mpc$^{3}$ {\it volume-limited} sample. 
The sample contains 47 galaxies which uniformally
sample the underlying galaxy population within the specified redshift,
magnitude and surface brightness limits  ($0.3 < z < 0.5$, 
$-21.3 < M_{B} < -13.7$ mags, $18.0 < \mu_{B} < 24.55$ mags/$\Box''$). 
We conclude:
(i) A luminosity-surface brightness relation exists for both 
the field and cluster galaxy populations, $M_{B} \approx [(1.5\pm0.2) \mu_{e} -
(50\pm2)]$,
(ii) Luminous low surface brightness galaxies account for
$<$ 10\% of the $L_{*}$ population,
(iii) Low luminosity low surface brightness galaxies outnumber Hubble types 
by a factor of $\sim 1.4$, however their space density is {\bf not} sufficient 
to explain the faint blue excess either by themselves or as faded remnants.
In terms of the local luminosity density and galaxy dynamical mass budget,
normal galaxies ({\it i.e.} Hubble tuning fork) contribute 88\% and 72\% 
respectively. This compares to 7\% and 12\% for dim galaxies 
and 5\% and 16\% for dwarf galaxies (within the above specified limits).
\end{abstract}




\section{Introduction}
The Hubble Deep Field \citep{bob96} and the many publications 
which have stemmed from these data are primarily associated with the
studies of faint galaxies \citep{spd98,fb98}, 
galaxy morphology \citep{ab96,sco96},
the evolution of luminous galaxies \citep{ab99} and 
the high redshift Universe in general. However the Hubble Deep Field (HDF) 
also provides us with the deepest insight of the local universe
probing into both the low surface brightness fog as well as the 
{\it intrinsically}
faint Universe. Since the original formulation of surface brightness selection 
effects and their potential impact on galaxy catalogs \citep{disney76}, the 
topic of low surface brightness galaxies has developed into a thriving area of 
scientific research \cite[see][for a recent review]{ib97}. However 
uncertainty still remains as to whether such systems represent a major or 
minor component of the galaxy population or more fundamentally the
mass budget. This uncertainty arises because of the
difficulty in firstly establishing a complete catalog of galaxies and 
secondly in obtaining the necessary distance measurements. Similarly the space 
density of intrinsically faint, or dwarf, galaxies is equally uncertain partly
because they are predominantly of low surface brightness, but more 
fundamentally because the 
volumes over which dwarf galaxies have been surveyed are small 
\cite[c.f.][]{spd96}. In general the space density of galaxies is poorly 
constrained. 
and can be quantified as a factor of $\sim 2$ at $L_{*}$ 
rising to a factor of $\sim 100$ at $0.01L_{*}$, and indeterminate faint wards 
(c.f. \citealp{z97} v's \citealp{lin96}).

What is required is an objective perspective of the entire local space
density of galaxies. That is a {\it volume}-limited sample over a wide and well
controlled range of surface brightness and intrinsic luminosity. 
Here we show how such a sample can be constructed from the Hubble Deep Field
utilizing photometric redshifts and quantify the contribution to the local 
luminosity density from normal, dim and dwarf galaxies.
We adopt a standard flat cosmology ($\Omega_{o}=1, \Lambda=0$), with
$H_{o}=75$km s$^{-1}$Mpc$^{-1}$ throughout.

\section{Constructing a Volume-limited sample}
In \citet{spd98} we combined the photometric redshift catalog
of \citet{fsoto98} with the morphological catalog
of \citet{sco96}. This catalog has now been extended a further magnitude to 
$I<27$ (now containing 675 galaxies), and updated to include quantitative 
measurements of the apparent half-light radii \cite[see][]{sco96}. 
Figure~\ref{fig1} shows a representation of this extended catalog by 
plotting each galaxy according to its redshift (x-axis) and its K-corrected
absolute magnitude (y-axis). These galaxies form a distribution which is 
bounded by the two appropriate apparent magnitude limits 
($B(F450W) > 19.5$ and $I(F814W) < 27$).
The reliability of photometric redshifts ($\Delta z \approx 0.1$) 
are discussed in \citet{hogg98}. In Figure~\ref{fig1} 
a horizontal line represents a sight-line across the past 10 Gyrs 
for a narrow absolute magnitude range and a vertical line represents a 
{\it volume}-limited sample at a specified redshift (within well defined 
magnitude and surface brightness limits). From Fig.~\ref{fig1} a 
{\it volume}-limited sample is 
{\it i.e.} any rectangle which lies within
the apparent magnitude limits. As we wish to explore the {\it local} 
galaxy population we shall concentrate on the low-z range generously 
redefined to $0.3 < z < 0.5$ ($326$ Mpc$^{3}$). 

The final sample contains 47 galaxies. Figure~\ref{fig2} 
shows these galaxies plotted according to their mean absolute surface 
brightness and their absolute B-band magnitude. The selection lines
shown on Fig.~\ref{fig2} are discussed in detail in \S 3. Note that
our chosen surface brightness measure is the intrinsic mean surface 
brightnesses ($\mu_{e}$) within the effective radius. 
The effective radius is derived from the measured azimuthally averaged 
half-light radii within the 25 I(F814W) mags per sq arcsecond isophote 
($r_{e}$).
Hence: 
\begin{equation}
\mu_{e} = m_{b}+2.5 log_{10}(2) + 2.5 log_{10} (\pi r_{e}^{2})
- 10 log_{10}(1+z)-K(z)
\end{equation}
The measured $r_{e}$ values are taken from \citet{sco96} and
the measured magnitudes and fitted K-corrections from \citet{fsoto98}.
This measurement of the surface brightness includes the bulge component
which explains why the values appear to be brighter than the more conventional
extrapolated central surface brightness - typically 21.7 mags per sq arcsec for
early-type disk systems \citep{f70}. 
Our motivation for using this 
measure of surface brightness is that no assumption of


\section{Selection Limits}
Having arbitrarily defined our redshift limits our absolute magnitude limits
are automatically set by the combination of these with the apparent 
magnitude constraints ($B > 19.5; I < 27$) as follows:
\begin{eqnarray}
M^{Upper}_{B} = & m^{bright}_{B}-D_{z_{LOW}}-K_{z_{LOW}} \\
M^{Lower}_{B} = & m^{faint}_{I}-D_{z_{HIGH}}-K_{z_{HIGH}}+(B-I)
\end{eqnarray}
Adopting $\Omega_{Total}=1, \Lambda=0$ and $H_{o}=75$kms$^{-1}$Mpc$^{-1}$, 
the distance moduli are: $D_{0.3}=40.5$, $D_{0.5}=41.7$. To be conservative we
adopt a K-correction suitable for a blue galaxy ($K_{0.3}=0.3$) at our upper 
limit and also a blue galaxy ($K_{0.5}=0.5$, $(B-I)=1.8$) for our lower limit
\citep{spd94}. This results in a {\it conservative} absolute 
magnitude range for completeness of: $-21.3 < M_{B} < -13.7$ - galaxies just
outside this range can be detected but not over the entire 326 Mpc$^{3}$
volume. The shaded regions on Fig.~\ref{fig1} 
show these selection limits over all redshift intervals
using a simplistic K-correction of K(z)=z. 
Note that the paucity of objects in the shaded region lends credence to
both the above magnitude limits and the reliability of photometric redshifts.

The upper surface brightness limit is defined by 
the point at which a galaxy with the lowest possible apparent magnitude is 
contained within a single resolution element (more luminous galaxies within a 
single pixel will have higher surface brightness measures).
For any absolute magnitude this will occur at the highest possible redshift 
({\it i.e.} z=0.5) hence the limit can be defined as follows:
\begin{equation}
\mu_{e}^{Min} < M_{B}+D_{z_{HIGH}}+2.5 log_{10}(2)+2.5 log_{10}(\pi (r_{e}^{Min})^{2}) - 10 log_{10}(1+z_{HIGH})
\end{equation}
Where $D_{0.5}$ is the distance modulus (41.7) and $r_e^{Min}$ is the 
resolution limit of the drizzled WFPC2 data ($r_{e}^{Max} =  0.04''$). 
Conversely the dimmest measurable surface brightness is defined by two limits. 
Firstly the fundamental isophotal detection limit at z=0.5, i.e.:
\begin{equation}
\mu^{B450W}_{limit} = \mu^{I814W}_{Iso}-10log_{10}(1+z_{HIGH})-K_{z_{HIGH}}+(B-I)
\end{equation}
and secondly by the maximum size at which an object can be detected, 
$r_e^{Max}$ (typically fixed by the f.o.v. or size of any smoothing 
filter). This results in a selection line as follows:
\begin{equation}
\mu_{e}^{Max} > M_{B}+D_{z_{LOW}}+2.5 log10(2)+2.5log(\pi (r_e^{Max})^{2})
-10 log_{10}(1+z_{LOW})
\end{equation}
Where $D_{0.3}=40.5$, and $r_e^{Max}=10''$.

\section{The Observed Bivariate Brightness Distribution}
On Figure~\ref{fig2} we map those objects with $0.3 < z < 0.5$ onto 
an effective absolute  surface brightness versus absolute magnitude plane.
Those galaxies which lie within these selection boundaries
represent the first {\it volume limited} census of the local galaxy population,
{\it i.e.} these 47 galaxies are a statistically uniform sample of the 
underlying galaxy population within these limits. In particular it surveys a 
larger volume for dim and dwarf galaxies than most 
\footnote{Only the LCRS \cite[c.f.][]{lin96} surveys a larger volume
for galaxies with $M_{B}=-14$ ($\sim700$Mpc$^{3}$), where peculiar motions and
survey incompleteness become significant due to the extreme closeness of the 
volume.} existing magnitude limited 
redshift surveys \citep[c.f.][]{spd96}, and extending to luminosities 
comparable to the brighter Local Group dwarfs \citep[c.f.][]{mateo}. 

\noindent
(1) {\bf The majority of galaxies lie along a magnitude-surface brightness 
relation ($M_{B} \approx [(1.5\pm0.2)\mu_{e} - (50\pm2)]$).} 
This is in very close agreement with that found for galaxies in the Virgo 
cluster ($M_{B} \propto 1.6\mu_{o} - K$, \citealp{b93}). 
This trend has recently been predicted from theoretical arguments \citep{jj} 
and through simulations of hierarchical merging 
\cite[e.g.][]{raul98, mo98}. Further work is required to allow 
direct comparisons between simulations - which in essence predict the mass 
versus angular momentum of the dark matter haloes - and observations - which 
determine the luminosity and surface brightness of the stellar population. 
It seems logical to suppose that the greater a galaxies
angular momentum the lower will be its surface brightness. Coupled with the
accepted correlation between luminosity and mass, the BBD may represent a key 
connection between easily obtained observables and fundamental physical 
properties.

\noindent
(2) {\bf Low surface brightness luminous galaxies are relatively rare.}
Low surface brightness galaxies have been postulated as a 
potentially grossly overlooked population \citep{disney76}, 
within which might be contained a substantial integrated mass \citep{ib97}. 
No such objects were identified in our 
326Mpc$^{3}$ volume. Unless our volume is unrepresentative the 
constraint is that luminous LSBGs (with $-21.3 < M_{B} < -18$ and 
$21.7 < \mu_{e} < 24.55$) are rare accounting for $<10$\% of the total $L_{*}$ 
population. Note that 21.7 mags per sq arcsec is adopted as the high/low 
surface brightness boundary, as this implies a system with a negligible
bulge component.
However we note that galaxies are known, such as Malin1 \citep{malin1}, 
which would not be detectable even in the HDF sight-line 
(due to both size and dimness).

\noindent
(3) {\bf At $0.3 < z < 0.5$ Dwarf galaxies are more numerous than giant 
galaxies}. Dwarf galaxies have been proposed as an explanation 
to the faint blue galaxy problem \cite[see][for a review]{ellis97}
either by postulating a dense inert population of dwarfs \citep[e.g.][]{spd94}
or through the recent fading of starbursting dwarfs to low 
surface brightness limits \citep[e.g.][]{spd96,bf96}.
In these two cases it is required locally for the dwarf-to-giant ratio to
be $\sim 50$ if non-evolving \citep{spd94} or a factor of
$\sim 5$ if evolving \citep{pd95}. The level of dwarfs seen
in Fig.~\ref{fig1} (dwarf-to-giant ratio = 1.4) does not support either of 
these scenarios. This suggests that 
while dwarf galaxies may contribute in part to the faint blue excess they
cannot be wholly responsible and a luminous population at higher redshift is
also required \cite[see also][]{spd98}. This dwarf-to-giant ratio is 
consistent with recent measures of the global luminosity function
\cite[e.g.][]{z97}, but inconsistent with the high dwarf-to-giant ratio of 
$\sim 4$ seen locally over the same absolute magnitude range \citep{kk99}. 
This may be due to the local volume being non-representative or alternatively
due to the greater scrutiny of the local environment
(i.e. even lower optical surface brightness limits and radio surveys).

\section{The Luminosity Density of the Universe}
Having constructed a volume-limited sample it is trivial to derive the 
luminosity-density of the Universe, 
{\it i.e.} $j_{B}=\sum^{\mu_{e}=18}_{\mu_{e}=24.55}\sum^{M_{B}=-21.3}_{M_{B}=-13.7} 
10^{0.4(M_{\odot}-M)}L_{\odot}$.
This is a fundamental parameter useful for cosmological purposes and can be 
combined with the maximum observed
mass-to-light ratio to obtain an {\it upper limit} to $\Omega_{Matter}$ 
\citep{cberg97}. Table~\ref{table1} 
shows the results for our HDF volume-limited sample
for all galaxies and also for subdivisions into giant (i.e. classical Hubble 
types), dim (i.e. LSBG disks) and dwarf systems.
Of interest, is that the value itself is typically 3-4
times larger than that obtained from local surveys. This is most likely
a statistical variation due to the small volume surveyed. However it may
also be a reiteration of a classic problem of faint galaxy models, namely the 
steepness of the local galaxy counts \citep{spd95a,ron98,spd95b}.
Finally we can now address the long standing question as to the cosmological 
importance of low surface brightness and dwarf galaxies.
Table~\ref{table1} 
shows the luminosity density for subregions of the BBD which 
fit with our definitions of high surface brightness giants, 
low surface brightness giants and dwarf galaxies as indicated. 
Adopting an invariant mass-to-light ratio
these values for the luminosity density would correspond directly to the 
percentage contribution to the galaxy mass density. However studies of both
low surface brightness galaxies and dwarf galaxies suggest mass-to-light ratios
typically increase towards both lower luminosity and lower surface 
brightnesses \citep{deblok96,kk99}. 
More work is require in this area, however for the 
moment we consider the results presented by \citet{z95} based
on observed Tully-Fisher relationships for low surface brightness galaxies of
$\frac{M}{L} \propto \Sigma_{o}^{-\frac{1}{2}}$.
We normalize this expression to the data of \citet{deblok96},
to give an $\frac{M}{L} = 3$ for 
normal high surface brightness disks ($-21.3 < M_{B} < -13.7, \mu_{e}=20.0$). 
The percentages based on these numbers are also shown in the final two 
columns of Table~\ref{table1} which finally reflects the local 
cosmological significance of high surface brightness, low surface brightness
and dwarf galaxies. To first order we see that high surface brightness galaxies
dominate both the luminosity density and to a slightly lesser extent the mass 
budget. Nevertheless the contribution from both low surface brightness and 
dwarf galaxies is non-negligible and may be partly responsible for the 
variation seen in measures of the local luminosity function (c.f. \S 1).

\section{Conclusions}
We have demonstrated that a {\it volume}-limited sample over a wide and well 
defined magnitude and surface-brightness range can be constructed from the
Hubble Deep Field and hence any deep, high resolution multi-color imaging 
survey. The volume-limited data has been mapped onto the first bivariate 
brightness distribution (BBD) for field galaxies. A clear result is a strong
luminosity-surface brightness relationship similar to that reported in
the Virgo cluster. We advocate the possible potential of the BBD as a
meeting point between simulations and observations by noting the logical
connections between surface brightness \& angular momentum and luminosity
\& mass. If shown to be true the BBD may represent a new and powerful
tool with which to trace galaxy evolution and environmental dependencies. 
More specifically the measured BBD over the 326 Mpc$^{3}$ volume and valid
for $0.3 < z < 0.5$, $-21.3 < M_{B} < -13.7$, $18.0 < \mu_{B} < 24.55$
shows no luminous low surface brightness galaxies and only a modest dwarf
population.  Hence we conclude that
no evidence is seen for a ``missing'' local population (within 
the above specified limits). 
In terms of the luminosity density and mass density we conclude that locally
high surface brightness giant galaxies dominate both the luminosity density 
(88 \%) and the galaxy contribution to the mass density of the local universe 
(72 \%). If a hierarchical model of galaxy formation is correct then these 
contributions will decrease with redshift. Such data will soon be
attainable with the {\it Hubble Space Telescope} Advance Camera and the 
{\it New Generation Hubble Space Telescope}.

\acknowledgments
Thanks to Alberto Fern\'andez-Soto and Steve Odewahn for help in assembling
the HDF catalog and to Bob Williams and the HDF team for the HDF
initiative.




\clearpage



\figcaption[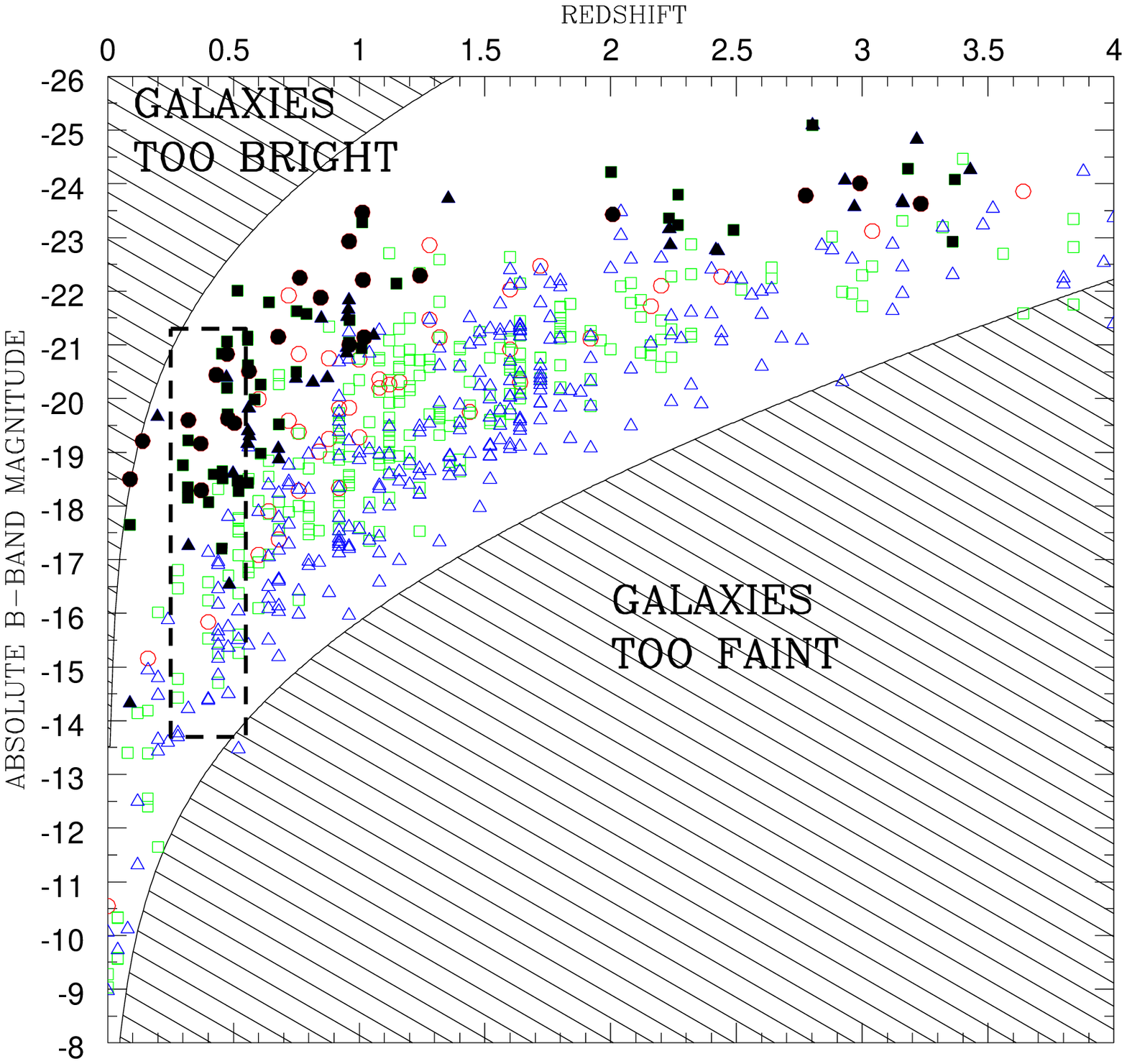]{
Galaxies with $I < 27$ from the Hubble Deep Field positioned
according to their absolute magnitude ($M_{B}$) and redshift (z). The shaded
region indicates the magnitude selection boundaries as defined in \S 3.1.
Symbols are: ellipticals (circles), spirals (asteriks) and irregulars
(triangles). Open symbols indicate photometric redshifts and solid
symbole spectroscopic redshifts. \label{fig1}}

\figcaption[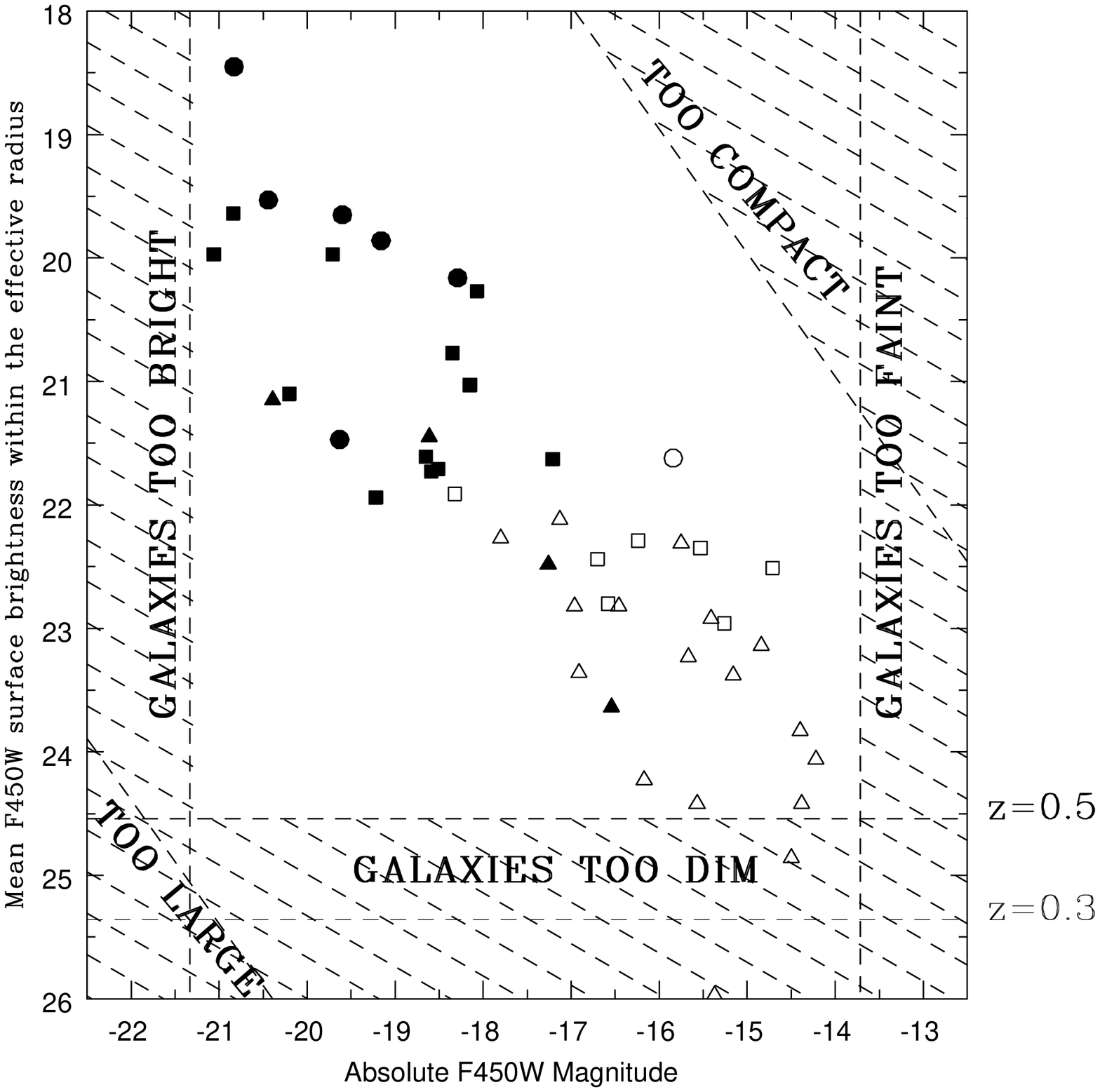]{
The bivariate brightness distribution for the HDF volume limited
sample and including various selection lines as indicated by the dashed lines. 
Symbols indicate morphological
classifications: E/S0 (open circles), Sabcs (asterisks), Sd/Irr (triangles)
\label{fig2}}





\clearpage


\begin{deluxetable}{lccccccc}
\tablecaption{The contribution of various galaxy generalizations to the 
luminosity density and the dynamical galaxy mass budget within the limits 
shown. \label{table1}}
\tablewidth{0pt}
\hspace{-2.0cm}
\tablehead{
\colhead{Galaxy} & \colhead{$M_{B}$}   &
\colhead{$\mu_{e}$} &
\multicolumn{2}{c}{$j$} & 
\multicolumn{2}{c}{$\rho_{M}^{Galaxies}$} \\
\colhead{Class} & \colhead{(mags)}   &
\colhead{(mags/$\Box''$)} &
\colhead{$(10^{8}L_{\odot}$Mpc$^{-3}$)}  & \colhead{(\%)} & 
\colhead{($10^{-28}$kg m$^{-3}$)} & \colhead{(\%)}
}
\startdata

All& $-21.3<M_{B}<-13.7$& $18.0<\mu_{e}<24.55$ & $8.2\pm1.2 $ & 100\% & $2.5\pm0.3$ & 100\% \\
Normal & $-21.3<M_{B}<-18$    & $18.0<\mu_{e}<21.7$ & $7.2\pm1.8$ & 88\% & $1.8\pm0.4$ & 72\% \\
Dim    & $-12.3<M_{B}<-18$    & $21.7<\mu_{e}<24.55$ & $0.6\pm0.3$ & 7\%  & $0.3\pm0.15$ & 12\% \\
Dwarf  & $-18<M_{B}<-13.7$ & $18.0<\mu_{e}<24.55$ & $0.4\pm0.1$ & 5 \% & $0.4\pm0.1$ & 16\% \\
 \enddata


\end{deluxetable}

\plotone{spd_hdf_figure1.ps}

\plotone{spd_hdf_figure2.ps}

\end{document}